\documentclass{aa}  
\usepackage{ulem}
\usepackage{graphicx}
\usepackage{txfonts}
\usepackage[pdftex,breaklinks,colorlinks,citecolor=blue,linkcolor=blue]{hyperref}
\usepackage[colorinlistoftodos]{todonotes}

\usepackage{subfig}
\usepackage{siunitx}
\usepackage{adjustbox}

\usepackage{color, colortbl}

\newcommand{\aminvsg}{$a_{\mathrm{min,\,a-C}}$}
\newcommand{\abvsg}{M$_{\mathrm{a-C}}$/M$_{\mathrm{H}}$}

\newcommand{\G}{$G_{0}$}

\newcommand{\qaf}{$q_{\mathrm{AF}}$}
\newcommand{\qpah}{$q_{\mathrm{PAH}}$}

\begin{document}

   \title{Influence of the nano-grain depletion in photon-dominated regions}

   \subtitle{Application to the gas physics and chemistry in the Horsehead}

   \author{T. Schirmer.
          \inst{1}
          E. Habart\inst{1}
          \and
          N. Ysard\inst{1}
          \and 
          E. Bron\inst{2}
          \and 
          J. Le Bourlot\inst{2}
          \and
          L. Verstraete\inst{1}
          \and 
          A. Abergel\inst{1}
          \and 
          A. P. Jones\inst{1}
          \and 
          E. Roueff\inst{2}
          \and 
          F. Le Petit\inst{2}
          }

   \institute{Université Paris-Saclay, CNRS,  Institut d'astrophysique spatiale, 91405, Orsay, France \\
   \email{thiebaut.schirmer@ias.u-psud.fr}
   \and 
   LERMA, Observatoire de Paris, PSL Research University, CNRS, Sorbonne Universités, 92190 Meudon, France.
             }
   \date{Received 26 February 2021; accepted 12 April 2021}

 
  \abstract
   {The large disparity in physical conditions from the diffuse interstellar medium (ISM) to denser clouds such as photon-dominated regions (PDRs) triggers an evolution of the dust properties (i.e. composition, size, and shape). The gas physics and chemistry are tightly connected to these dust properties and are therefore affected by dust evolution and especially the nano-grain depletion in the outer irradiated part of PDRs.}
   {We highlight the influence of nano-grain depletion on the gas physics and chemistry in the Horsehead nebula, a prototypical PDR.}
   {We used a model for atomic and molecular gas in PDRs, the Meudon PDR code, using diffuse ISM-like dust and Horsehead-like dust to study the influence of nano-grain depletion on the gas physics and chemistry, focusing on the impact on photoelectric heating and H2 formation and, therefore, on the H$_2$ gas lines.}
   {We find that nano-grain depletion in the Horsehead strongly affects gas heating through the photoelectric effect and thus the gas temperature and the H$_2$ formation, hence the H~$\rightarrow$~H$_2$ position. Consequently, the first four pure rotational lines of H$_2$ (e.g. 0-0 S(0), S(1), S(2), and S(3)) vary by a factor of 2 to 14. The 0-0 S(3) line that is often underestimated in models is underestimated even more when taking nano-grain depletion into account due to the decrease in gas heating through the photoelectric effect. This strongly suggests that our understanding of the excitation of H$_2$ and/or of heating processes in the Horsehead, and more generally in PDRs, is still incomplete.}
   {Nano-grain depletion in the outer part of the Horsehead has a strong influence on several gas tracers that will be prominent in JWST observations of irradiated clouds. We therefore need to take this depletion into account  in order to improve our understanding of the Horsehead, and more generally PDRs, and to contribute to the optimal scientific return of the mission.}
   {}


   \maketitle

\section{Introduction}\label{sect:introduction}

Interstellar dust plays an important role in numerous physical and chemical 
processes in the interstellar medium (ISM) and most particularly in photon-dominated regions (PDRs). In the outer irradiated part of PDRs, gas heating is mainly due to the photoelectric effect on dust grains \citep[e.g.][]{bakes_photoelectric_1994} and the formation of molecular hydrogen H$_2$ \citep[e.g.][]{le_bourlot_surface_2012, bron_stochastic_2014, jones_h_2015} happens on dust surfaces. The efficiency of these processes strongly depends on the dust physical properties (size, composition, and shape) and must therefore be constrained to understand PDR observations, such as the H$_2$ gas lines.

With the aim of modelling PDRs, several models have been developed \citep[see][for a comparative study of these codes]{rollig_photon_2007}, mostly using grains with properties similar to those of the diffuse ISM \citep[e.g.][]{mathis_size_1977, weingartner_dust_2001, draine_infrared_2007, compiegne_global_2011, jones_evolution_2013}. However, it has been shown that dust evolves from diffuse to dense regions \citep[e.g.][]{flagey_evidence_2009-1, ysard_variation_2013, ysard_mantle_2016, juvela_dust_2020, saajasto_multi-wavelength_2021} and most specifically in PDRs \citep[e.g.][]{berne_analysis_2007, abergel_evolution_2010, pilleri_evaporating_2012, arab_evolution_2012, van_de_putte_evidence_2019, schirmer_dust_2020}. Based on the Spitzer and Herschel observation together with the THEMIS dust model \citep{jones_evolution_2013,jones_global_2017} and the radiative transfer code SOC \citep{juvela_soc_2019}, \cite{schirmer_dust_2020} constrained the dust properties in the Horsehead nebula. They found that in the outer irradiated part of the Horsehead (i.e. below a maximum depth of 0.05 pc from the edge, where H$_2$ and nano-grains emit), the nano-grain dust-to-gas ratio is 6-10 times lower and the minimum size of these grains is 2-2.25 times larger than in the diffuse ISM. In the inner part of the Horsehead, they found that  grains most likely consist of multi-compositional mantled aggregates. Thus, the efficiency of physical and chemical processes in PDRs is affected. Using the Meudon PDR code \citep{le_petit_model_2006}, \cite{goicoechea_penetration_2007} showed that grain growth with depth inside a PDR affects the penetration of the UV flux and therefore the physical and chemical structures inside this region. However, they did not study the influence of grain growth on the efficiency of specific physical and chemical mechanisms such as heating through the photoelectric effect on dust or the H$_2$ formation.
In this paper, we propose a complementary study.

We show the influence of the nano-grain depletion in the outer irradiated part of the Horsehead \citep[][]{schirmer_dust_2020} on the physics and chemistry of this region using the Meudon PDR code \citep{le_petit_model_2006}. The paper is organised as follows. In Sect.\,\ref{sect:models_tools}, we present the Meudon PDR code. We also describe how we tuned the dust model in the Meudon PDR code in order to reproduce the observed nano-grain depletion in the Horsehead using the THEMIS dust model \citep{jones_evolution_2013, jones_global_2017}. In Sect.\,\ref{sect:results}, we present the influence of nano-grain depletion on the gas temperature, the H~$\rightarrow$~H$_2$ transition, and several H$_2$ gas tracers. Finally, we summarise and conclude in Sect.\,\ref{sect:summary_and_conclusion}.

\section{Models and tools}\label{sect:models_tools}

In order to study the influence of nano-grain depletion on the gas physics and chemistry in the outer irradiated region of the Horsehead, we used a model for atomic and molecular gas in PDRs, the Meudon PDR code. First, we present this model.\ Second, we describe how we tuned the dust parameters to reproduce, at first order, the observed nano-grain depletion in the Horsehead.

\subsection{The Meudon PDR code}\label{sect:sect:Meudon_PDR_code}

The Meudon PDR code\footnote{The Meudon PDR code is available here: \href{ ism.obspm.fr}{ ism.obspm.fr}} \citep{le_petit_model_2006} is a model for atomic and molecular gas in irradiated interstellar clouds. It uses a 1D parallel plane geometry in aid of a thorough description of the physical and chemical processes at work inside PDRs. Up to 234 species are included and involved in a chemical network that is described in the chemistry file\footnote{We use the version 1.5.4: \href{https://ism.obspm.fr/pdr_download.html}{https://ism.obspm.fr/pdr$\_$download.html}} provided with the code. The dust model used in this code is based on an MRN distribution \citep{mathis_size_1977} for graphites and silicates with a possibility to add a log-normal distribution of PAHs as in \cite{draine_infrared_2007} (see Fig.\,\ref{fig:mass_distri}). The description of these grains is important as they are involved in  i) the radiative transfer through their extinction \citep{goicoechea_penetration_2007}; ii) the chemistry of H$_{2}$ and other species by way of their catalyst role \citep{le_bourlot_surface_2012, bron_surface_2014}; and iii) the thermal balance with both the photoelectric effect \citep{bakes_photoelectric_1994, weingartner_photoelectric_2001} and the gas-grain coupling.

\subsection{Parametrisation of nano-grain depletion in the Horsehead with the Meudon PDR code}\label{sect:sect:Meudon_PDR_code}

Based on \textit{Herschel} and \textit{Spitzer} observations of the Horsehead together with the THEMIS\footnote{THEMIS is available here: \href{https://www.ias.u-psud.fr/themis/index.html}{https://www.ias.u-psud.fr/themis/}} dust model \citep{jones_evolution_2013,jones_global_2017}, \cite{schirmer_dust_2020} show a strong depletion of nano-grains in the irradiated outer part of this PDR.\ The nano-grain dust-to-gas ratio is 6-10 times lower and the minimum size of these grains is 2-2.25 times larger than in the diffuse ISM. 

\begin{figure}[h]
\centering
\includegraphics[width=0.5\textwidth, trim={0 0cm 0cm 0cm},clip]{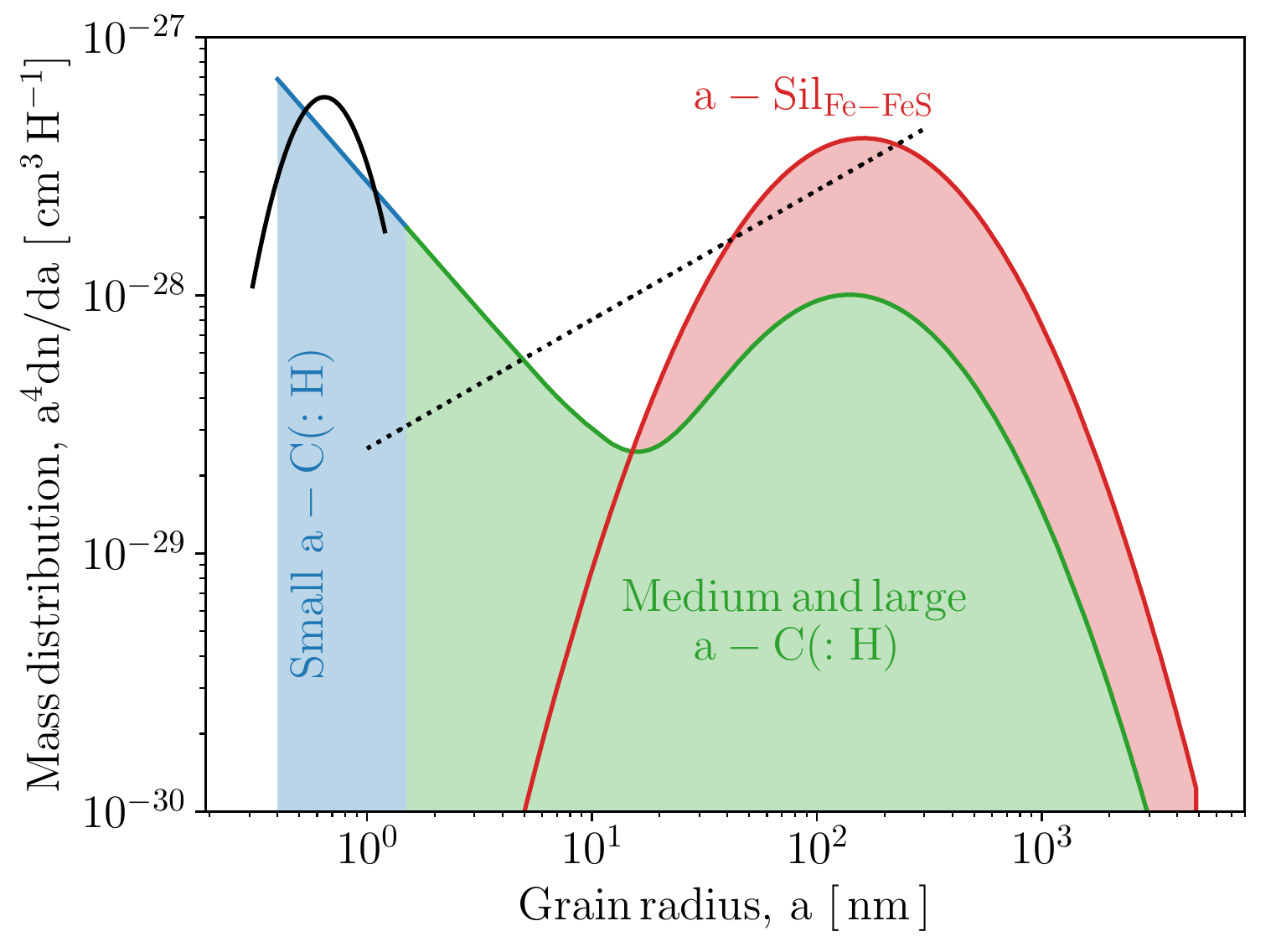}
    \caption{Mass distribution for small a-C(:H) in blue, medium and large a-C(:H) in green, and silicate grains in red. These sub-components composed the THEMIS dust model for the diffuse ISM. The PAH mass distribution used in the Meudon PDR code is shown with a black line. The MRN distribution of graphite and silicates (70 $\%$ graphite and 30 $\%$ silicate) used in the Meudon PDR code is shown with a dotted line.}
    \label{fig:mass_distri}
\end{figure}

In order to explore, qualitatively, the effects of this nano-grain depletion on the gas using the Meudon PDR code, we needed to tune parameters of the dust model used in the Meudon PDR code with the aim of consistently reproducing the grain depletion within the THEMIS framework. To this end, we tuned the parameter \qpah~which corresponds to the PAH-to-dust mass ratio. This parameter is equivalent to the parameter \qaf~\citep[see][]{galliano_nearby_2021} which is the mass fraction of aromatic-feature-carrying grains in THEMIS, which are the small a-C(:H) grains in Fig.\,\ref{fig:mass_distri}. In the diffuse ISM, \qpah~=~$4.6~\%$ \citep[see][]{draine_infrared_2007}, whereas \qaf~= $17~\%$~\citep[see][]{galliano_nearby_2021}. This difference can be explained because the THEMIS model does not contain PAHs, but a-C(:H) material that is composed of less aromatic bonds per dust mass than PAHs. It is therefore necessary to consider a higher mass of a-C(:H) to reproduce the same aromatic emission as with PAHs. As dust evolves in the ISM, these two ratios evolve as well and are linked by the following relation:

\begin{equation}
    \label{eq:qpah_qaf}
    q_{\mathrm{PAH}} = \frac{4.6}{17} \times q_{\mathrm{AF}} = 0.27 \times q_{\mathrm{AF}}
.\end{equation}

\begin{figure}[h]
\centering
\includegraphics[width=0.5\textwidth, trim={0 0cm 0cm 0cm},clip]{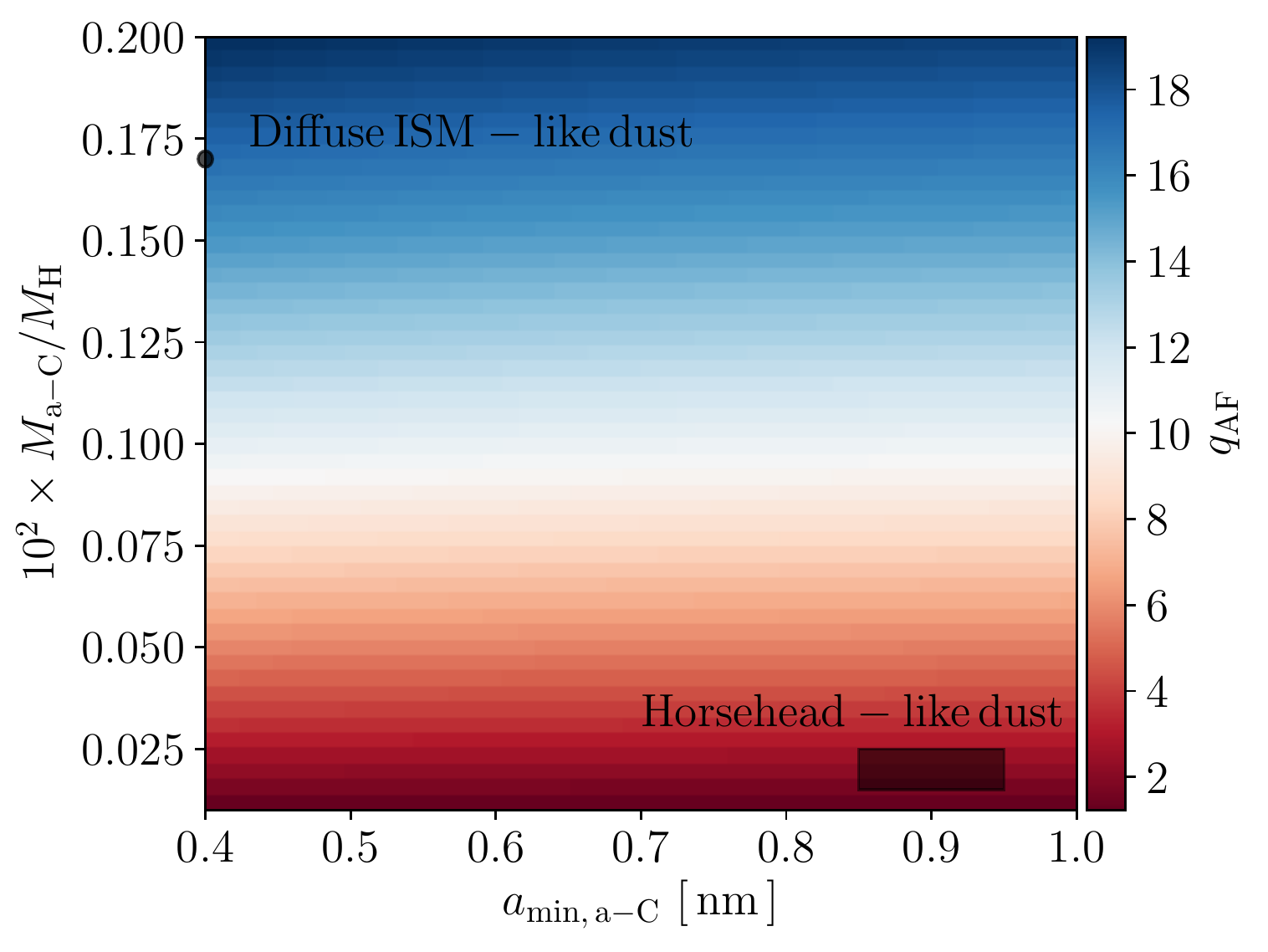}
    \caption{Mass fraction \qaf~of aromatic-feature-carrying grains in THEMIS as a function of the nano-grain dust-to-gas ratio \abvsg~and the nano-grain minimum size \aminvsg. The grey dot corresponds to the dust properties of the diffuse ISM-like dust (\abvsg~= $0.17\times 10^{-2}$ and \aminvsg~= $0.4$ nm). The grey rectangle corresponds to the dust properties in the irradiated outer part of the Horsehead.}
    \label{fig:q_AF}
\end{figure}

Figure \ref{fig:q_AF} shows \qaf~as a function of both the nano-grain dust-to-gas ratio \abvsg~and the nano-grain minimum size \aminvsg. First and foremost, \qaf~decreases from $17~\%$ in the diffuse-ISM to $\sim 2~\%$ in the Horsehead. We also note that \qaf~barely changes with \aminvsg~as an increase in this parameter implies a redistribution of the nano-grain mass towards larger nano-grains \citep[see Fig.\,4 in][]{schirmer_dust_2020} that are mostly smaller than $1.5$ nm. From now onwards, a decrease in \qaf~is equivalent to an increase in the depletion of nano-grains and we assume two values of reference that are \qaf~= 17\,\% in the diffuse ISM and \qaf~=~2\,\% in the outer irradiated part of the Horsehead.

As it is not currently possible to use dust properties varying with $A_\mathrm{V}$ in the Meudon PDR code, we only focus on a smaller region of the outer irradiated part of the Horsehead (i.e. below a maximum depth of 0.025 pc from the edge), where there are no aggregates. This region corresponds to where the H~$\rightarrow$~H$_2$ transition happens and where most of the H$_2$ lines emit. For the sake of simplicity, we used the same extinction curve regardless of \qaf~(and therefore a constant $R_{\mathrm{V}}$ = 3.1). This means that we did not take the influence of the nano-grain depletion on the radiative transfer into account. This assumption is discussed further in Sect.\,\ref{sect:sect:assumption}.

\section{Results}\label{sect:results}

We now present the results of our gas modelling in the Horsehead using the Meudon PDR code for different values of \qaf. Firstly, we describe how we modelled the Horsehead with the Meudon PDR code (see Sect.\,\ref{sect:sect:HH}); and secondly, we present the influence of dust evolution through the parameter \qaf~on the gas temperature (see Sect.\,\ref{sect:sect:heating_source}), the chemistry (see Sect.\,\ref{sect:sect:chemistry}), and gas tracers (see Sect.\,\ref{sect:sect:gas_observables}).

\subsection{Horsehead modelling}\label{sect:sect:HH}
As it is possible to use a density profile in the Meudon PDR code and with the aim of being consistent with the study of \cite{schirmer_dust_2020}, we used the same density profile described in \cite{habart_density_2005} and the same radiation field whose intensity is about $G_0\sim 100$. The other parameters are defined in Table\,\ref{tab:PDR_model_parameters}. One may note that the standard ionisation rate $\zeta$ in the Meudon PDR code is $1 \times 10^{-17}$ s$^{-1}$, albeit \cite{rimmer_observing_2012} found a cosmic ray ionisation rate of $5 \times 10^{-15}$ s$^{-1}$ in the Horsehead. However, we show (see Appendix \ref{appendix:cosmic_ray}) that variations in $\zeta$ from $1 \times 10^{-17}$ s$^{-1}$ s$^{-1}$ to $5 \times 10^{-15}$ s$^{-1}$ result in changes in the H$_2$ gas tracers of the order of less than one percent. Our results are therefore unaffected by variations in $\zeta$ and we set this rate to $1 \times 10^{-17}$ s$^{-1}$. From this point onwards, we focus on a smaller region of the outer irradiated part of the Horsehead where the H $\rightarrow$ H$_{2}$ transition happens and where most of the H$_2$ gas tracers emit (i.e. $d<0.03$ or $d<2.25\arcsec$, see Fig.\,\ref{fig:temp_q_pah} top panel). In order to study the influence of dust evolution on the gas physics and chemistry, we modified the parameter \qpah~from 0 \% to 6 \% (\qaf~varying from 0 \% to 22 \%, see Eq.\,\eqref{eq:qpah_qaf}) on a linear grid composed of 35 points.

\begin{table*}[h]
    \centering
    \begin{tabular}{lll}
        \hline
        \hline
        Parameter & Value & Comment \\
         \hline
         $G_0$ & 100 & radiation field in Habbing unit \\
        $\zeta$ & $5\times 10^{-17}$ s$^{-1}$ & cosmic rays ionisation rate\\
        $v_{\mathrm{turb}}$ &  3 km\,s$^{-1}$ & turbulent velocity \\
        $P_{\mathrm{th}}$ & $3\times 10^{6}$ K\,cm$^{-3}$ & pressure \\
        $R_{V}$ & 3.1 & optical total-to-selective extinction\\
        $a_{\mathrm{min}}$ & 1 nm & grain minimum radius \\
        $a_{\mathrm{max}}$ & 300 nm & grain maximum radius \\
        $\alpha$ & 3.5 & size distribution power-law index \\
         \hline
    \end{tabular}
    \caption{PDR model parameters.}
    \label{tab:PDR_model_parameters}
\end{table*}

\subsection{Thermal balance}\label{sect:sect:heating_source}

\begin{figure}[h]
\centering
\includegraphics[width=0.5\textwidth, trim={0 0cm 0cm 0cm},clip]{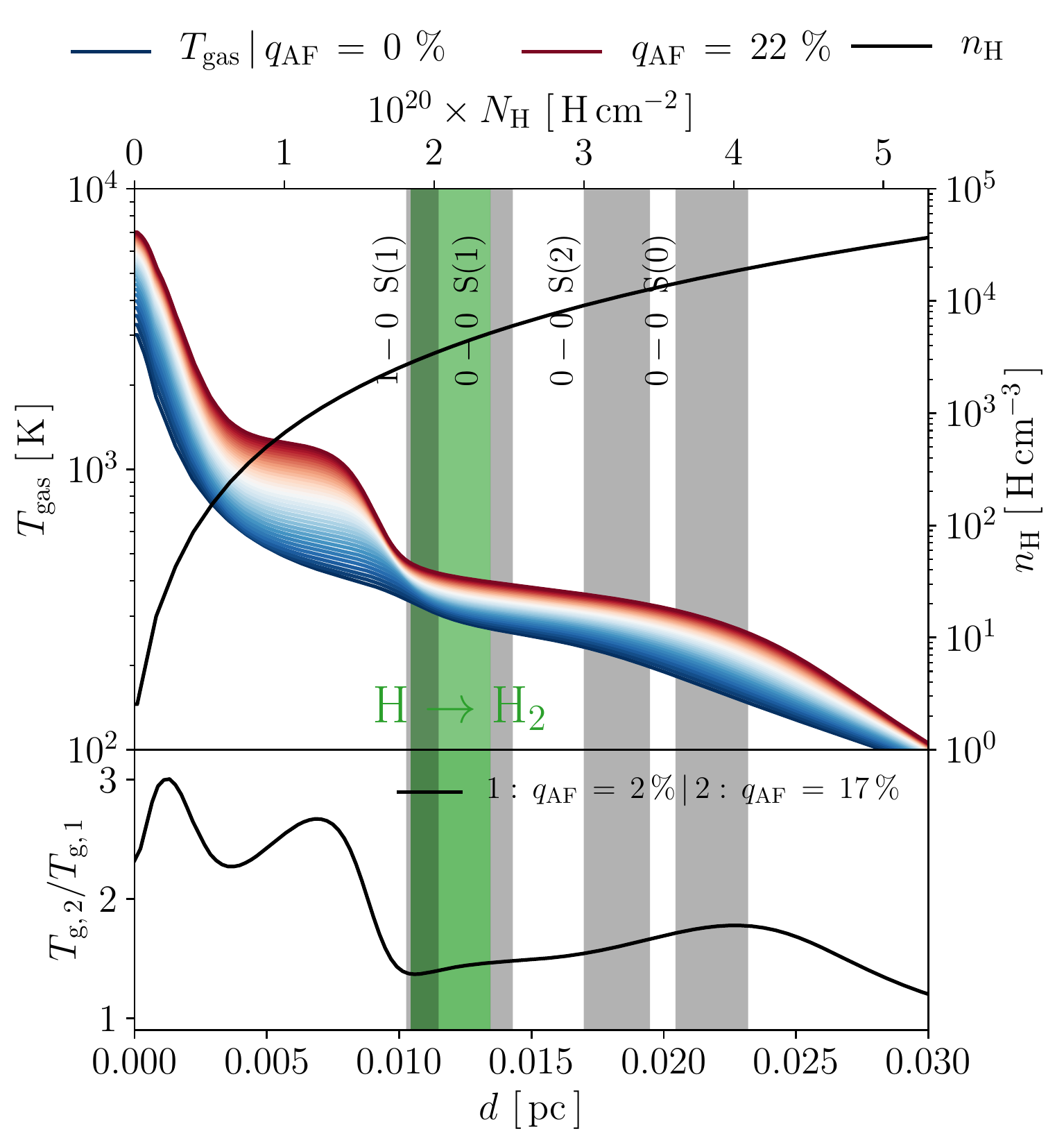}
    
    \caption{\textit{Top:} Gas temperature across the PDR for \qaf~varying from $0~\%$ (blue line) to $22~\%$ (red line). The density profile is shown with a black line. \textit{Bottom:} Ratio between the gas temperature across the PDR using the diffuse ISM-like dust (\qaf~= 17 \%) and the Horsehead-like dust (\qaf~= 2 \%). The green stripe corresponds to the position of the H $\rightarrow$ H$_2$ transition (see Fig.\,\ref{fig:position_transition}) for \qaf~varying from 2 $\%$ (Horsehead-like dust) to 17 $\%$ (diffuse ISM-like dust). The four grey bands successively correspond\ to the position of the maximum emission of the H$_2$ 1-0 S(1) line (see Fig.\,\ref{fig:1_0_S_1}), the H$_2$ 0-0 S(1) line (see Fig.\,\ref{fig:0_0_S_1}), the H$_2$ 0-0 S(2) line (see Fig.\,\ref{fig:0_0_S_2}), and the H$_2$ 0-0 S(0) line (see Fig.\,\ref{fig:0_0_S_0}) for 2 $\leq$ \qaf~ $\leq$ 17 \%.}
    \label{fig:temp_q_pah}
\end{figure}

Figure \ref{fig:temp_q_pah} (top panel) shows the gas temperature across the Horsehead for \qaf~varying from 0 $\%$ to 22 $\%$, as well as both the position of the H~$\rightarrow$~H$_2$ transition and the position of the maximum of several H$_2$ lines (0-0 S(0), S(1), S(2), and 1-0 S(1)). The density at the H~$\rightarrow$~H$_2$ transition is about $4-7\times 10^{3}$ H\,cm$^{-3}$ and $T_{\mathrm{gas}}=3-4\times 10^{2}$ K, hence the thermal pressure\footnote{$P_{\mathrm{th}}=n_{\mathrm{H}}\times T_{\mathrm{gas}}$} is about $P_{\mathrm{th}}=1.2-2.8 \times 10^{6}$ K\,cm$^{-3}$. In addition, \cite{joblin_structure_2018} find the following relation between the thermal pressure and \G~in PDRs: $P_{\mathrm{th}}/G_0=5\times 10^{3}-8\times 10^{4}$ K\,cm$^{-3}$, which when applied to the Horsehead provides us with the following range of expected thermal pressure $P_{\mathrm{th}}=5\times 10^{5}-8\times 10^{6}$ K\,cm$^{-3}$. Our model is therefore consistent with \cite{joblin_structure_2018}.

\begin{figure}[h]
\centering
\includegraphics[width=0.5\textwidth, trim={0 0cm 0cm 0cm},clip]{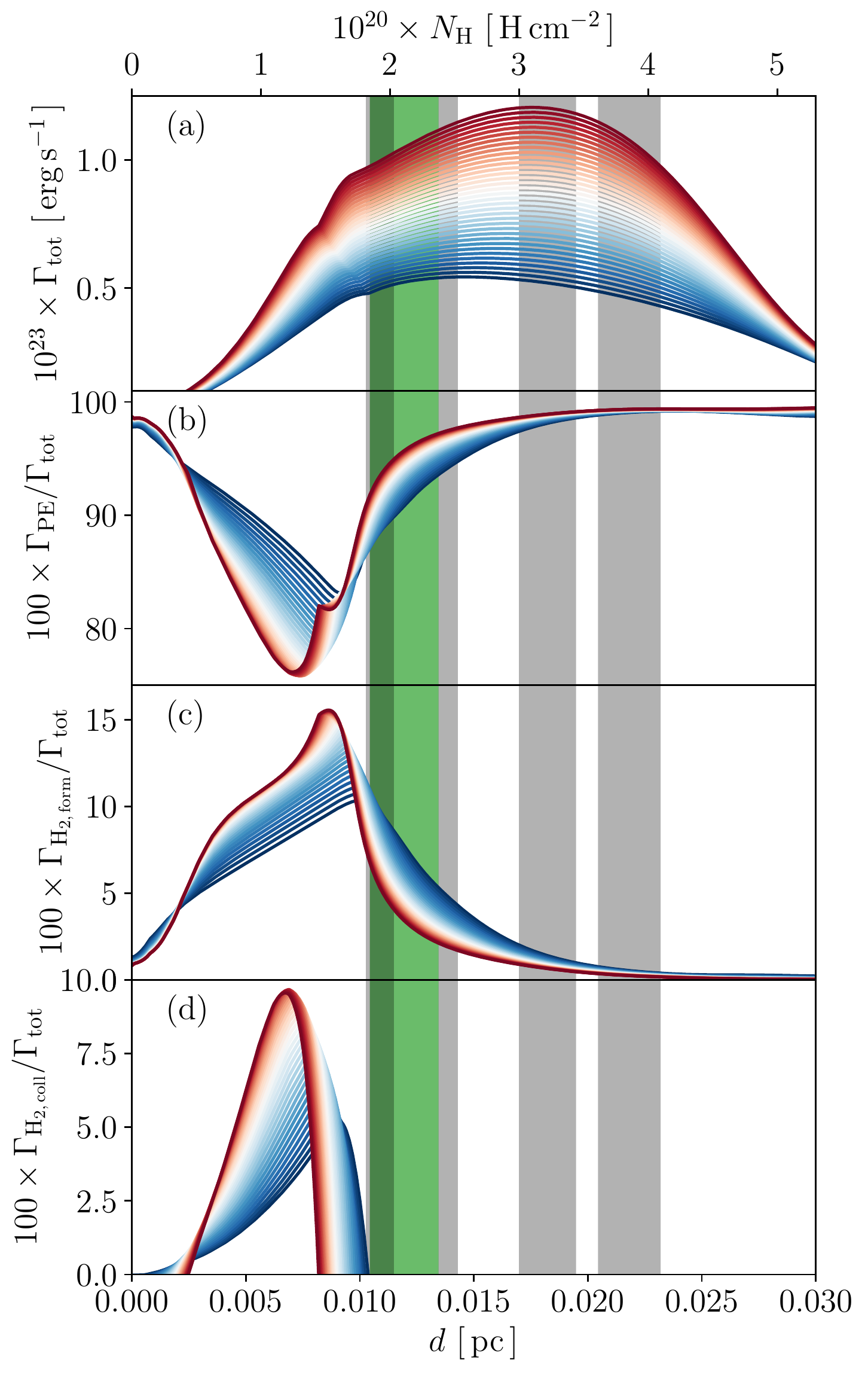}
    \caption{\textit{Panel a}: Total gas heating rate across the PDR for \qaf~varying from $0~\%$ (blue line) to $22~\%$ (red line). \textit{Panel b}: Contribution of the gas heating through the photo-electric effect on dust to the total gas heating. \textit{Panel c}: Contribution of the gas heating through the H$_2$ formation  to the total gas heating. \textit{Panel d}: Contribution of the gas heating through H$_2$ collisional deexcitation to the total gas heating. The meaning of the green and grey stripes is explained in Fig.\,\ref{fig:temp_q_pah}.}
    \label{fig:heating_q_pah}
\end{figure}

Regardless of the depth inside the Horsehead, the gas temperature decreases with \qaf~from a factor of 1.5 to 3 (see Fig.\,\ref{fig:temp_q_pah}, bottom panel) when using Horsehead-like dust (i.e. \qaf~= 2 \%) instead of diffuse ISM-like dust (i.e. \qaf~= 17 \%). These variations in the gas temperature with \qaf~are due to variations in the gas total heating with \qaf~(see Fig.\,\ref{fig:heating_q_pah}, panel \textit{a}). Furthermore, the gas heating through the photoelectric effect on dust contributes at least to 80 \% of the total gas heating, hence we can consider that dust heating in the outer irradiated part of the Horsehead is dominated by this process (see Fig.\,\ref{fig:heating_q_pah}, panel \textit{b}). As \cite{bakes_photoelectric_1994} showed that the photoelectric effect is only efficient on nano-grains and because a decrease in \qaf~implies a decrease in this grain population, the heating via the photoelectric effect on dust grains decreases as well and therefore explains the decrease in the gas temperature with \qaf.

Even if the other gas heating processes are marginal compared to the photoelectric effect on dust, the heating through the H$_2$ formation can contribute up to 15 \% of the total gas heating (see Fig.\,\ref{fig:heating_q_pah}, panel \textit{c}) and up to 10 \% through the H$_2$ collisional deexcitation (see Fig.\,\ref{fig:heating_q_pah}, panel \textit{d}). Regarding the gas heating due to the H$_2$ formation, it increases with \qaf~because the H$_2$ formation increases with \qaf~(see Fig.\,\ref{fig:formation_H2}). As the UV radiation field is high at the Horsehead edge, H$_2$ is efficiently pumped by external UV radiation. This process is usually followed by collisional deexcitations that can heat the gas. However, a density of about $10^{3}$ H\,cm$^{-3}$ \citep{hollenbach_photodissociation_1999} must be reached to efficiently heat the gas through this process. In our PDR model, the density profile reaches this typical value where the UV radiation field is still high enough to efficiently pump H$_2$ (i.e. before the H $\rightarrow$ H$_2$ transition); however, since the density is not much higher than $10^{3}$ H\,cm$^{3}$ ($n_{\mathrm{H}}$< 2000 H\,cm$^{3}$), the efficiency of the heating through H$_2$ collisional deexcitations remains low.

\subsection{Chemistry}\label{sect:sect:chemistry}
\begin{figure}[h]   
\centering
\includegraphics[width=0.5\textwidth, trim={0 0cm 0cm 0cm},clip]{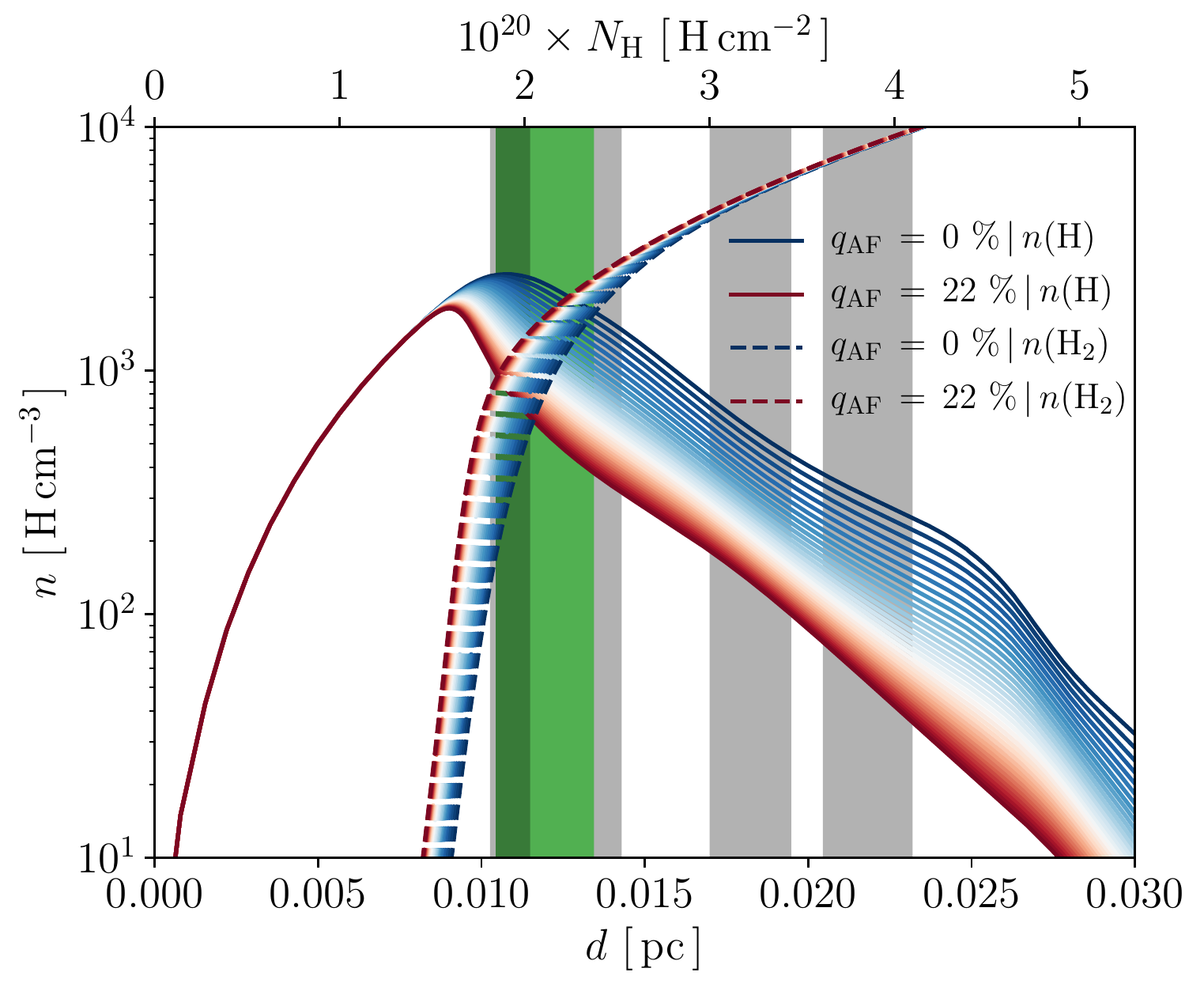}
    \caption{Density of atomic hydrogen (solid line) and molecular hydrogen (dashed line) across the PDR for \qaf~varying from $0~\%$ (blue line) to $22~\%$ (red line). The meaning of the green and grey stripes is explained in Fig.\,\ref{fig:temp_q_pah}.} 
    \label{fig:density_q_pah}
\end{figure}

\begin{figure}[h]
\centering
\includegraphics[width=0.5\textwidth, trim={0 0cm 0cm 0cm},clip]{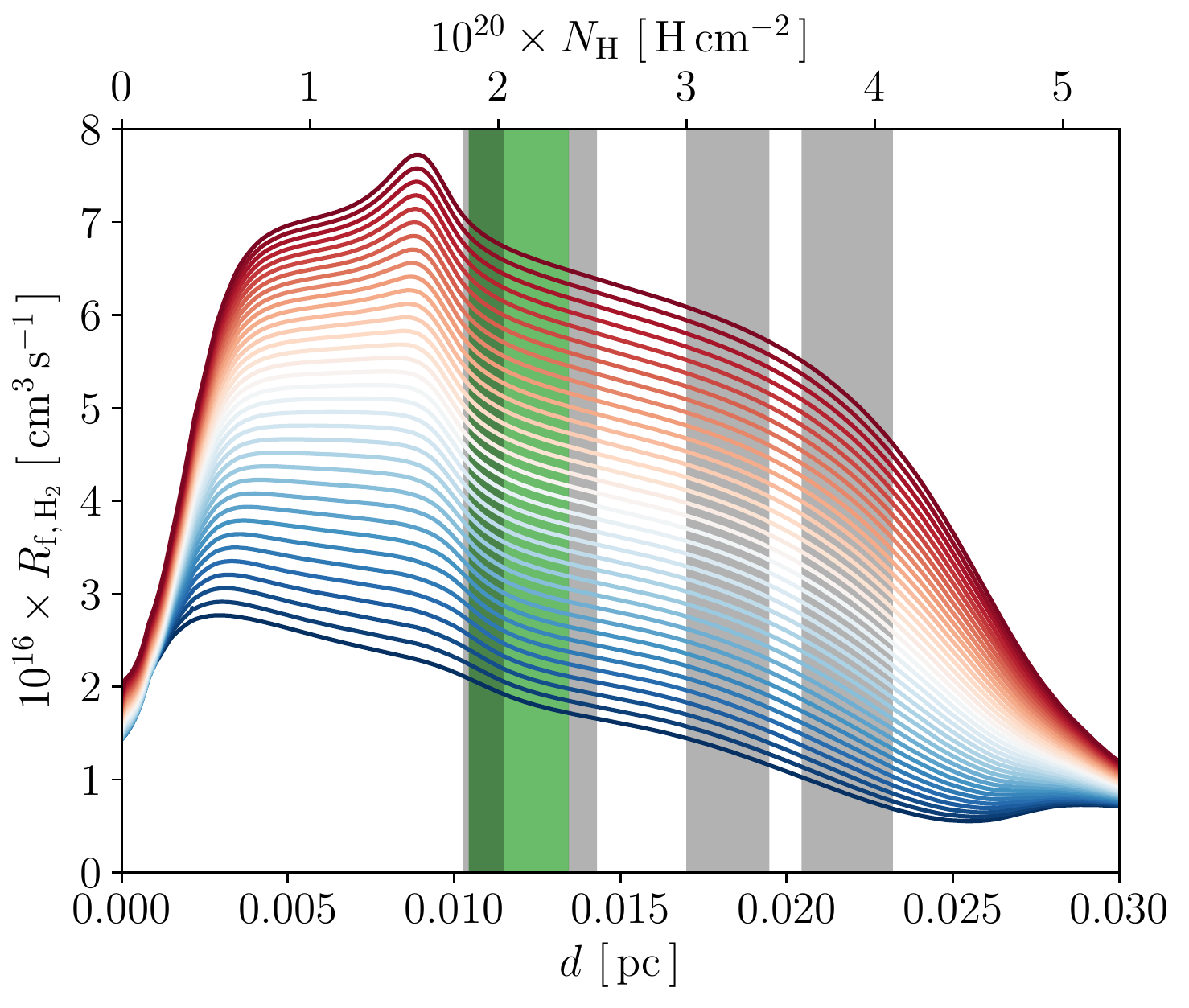}
    \caption{Total H$_2$ formation rate across the PDR for \qaf~varying from $0~\%$ (blue line) to $22~\%$ (red line). The meaning of the green and grey stripes is explained in Fig.\,\ref{fig:temp_q_pah}.}
    \label{fig:formation_H2}
\end{figure}

Figure \ref{fig:density_q_pah}
shows the density of atomic and molecular hydrogen across the Horsehead for \qaf~varying from 0~\% to 22~\%. Regardless of \qaf, the H$_2$ abundance increases with depth inside the Horsehead until it exceeds that of atomic hydrogen H. As the UV flux decreases with depth inside the Horsehead because of dust extinction and H$_2$ self-shielding, H$_2$ photodissociation decreases as well together with an increase in the H$_2$ formation (see Fig.\,\ref{fig:formation_H2}), leading to an increase in the H$_2$ abundance. 

\begin{figure}[h]   
\centering
\includegraphics[width=0.5\textwidth, trim={0 0cm 0cm 0cm},clip]{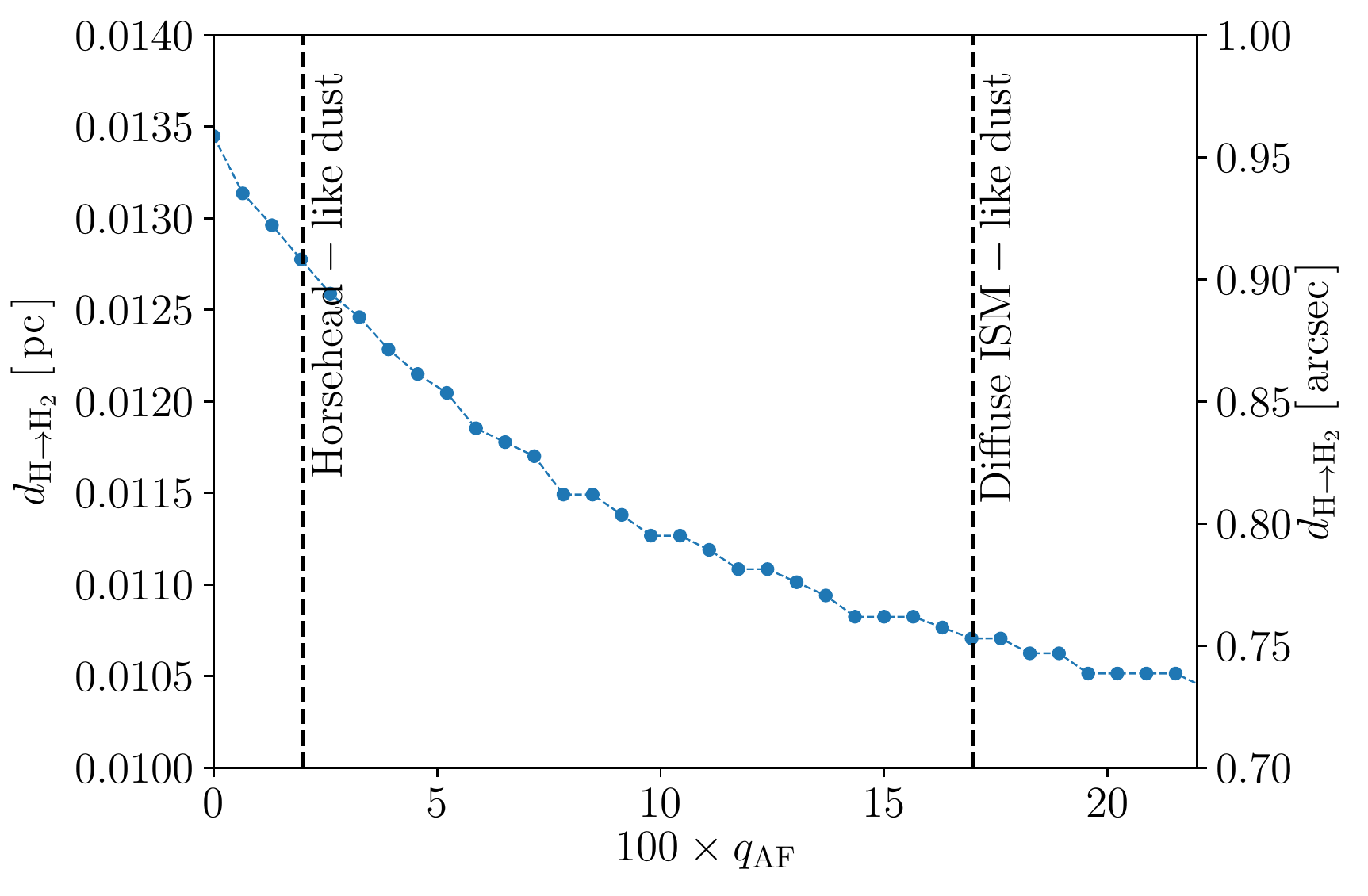}
    \caption{Position of the H $\rightarrow$ H$_2$ transition as a function of \qaf.} 
    \label{fig:position_transition}
\end{figure}

The position of the H~$\rightarrow$~H$_2$ transition also varies with \qaf. Figure \ref{fig:position_transition} shows the transition position as a function of \qaf. A decrease in \qaf~implies a shift of the H $\rightarrow$ H$_2$ towards the inner part of the Horsehead because such a decrease in \qaf~implies a decrease in the H$_2$ formation rate. Indeed, figure \ref{fig:formation_H2} shows the H$_2$ formation rate across the Horsehead for \qaf~varying from 0~\% to 22~\%. The molecular hydrogen H$_2$ mainly forms on the grain surface \citep[e.g.][]{hollenbach_surface_1971} and is modelled in the Meudon PDR code through two different mechanisms \citep{le_bourlot_surface_2012} that are the following:

\begin{description}
    \item[$\bullet$ \textbf{the Langmuir-Hinshelwood mechanism}] that consists of a physisorbed hydrogen atom that can migrate on the dust surface until it encounters another physisorbed hydrogen atom and therefore forms H$_2$. For high dust temperatures (i.e. typically in the outer irradiated part of PDRs), large grains are too hot and the hydrogen atoms are thermally desorbed from the dust surface before forming H$_2$. The efficiency of this process is therefore low in the outer irradiated part of PDRs. However, \cite{bron_surface_2014} show that this mechanism can be efficient on nano-grains as they are stochastically heated and therefore can stay in the temperature range long enough, allowing H$_2$ to form before the hydrogen atoms are thermally desorbed. When the hydrogen atoms are not thermally desorbed (i.e. when the temperature is low enough) and in the same time when they can migrate (i.e. when the temperature is high enough), the efficiency of this mechanism only depends on the dust total surface;
    \item[$\bullet$ \textbf{the Eley-Rideal mechanism}] that consists of a chemisorbed hydrogen atom on the dust surface which is impinged by a hydrogen atom of the gas phase thereby forming H$_2$. The efficiency of this mechanism is proportional on the dust total surface and sensitive to the gas temperature.  
\end{description}

In the version of the Meudon PDR code we used, the H$_2$ formation is described by the formalism of \cite{le_bourlot_surface_2012} and it does not contain the results of \cite{bron_surface_2014}. In our model, the Langmuir-Hinshelwood mechanism is therefore not efficient enough to form H$_2$ in the outer part of the Horsehead and it is marginal compared to the Eley-Rideal mechanism. The H$_2$ formation is thus proportional to the gas temperature and the total dust surface. As both of these quantities decrease with \qaf, the H$_2$ formation decreases with \qaf~as well. We would have the same effect if the formalism of \cite{bron_surface_2014} would have been included in the Meudon PDR code as the efficiency of the Langmuir Hinshelwood mechanism is proportional to the total dust surface. Also, regardless of \qaf~and the PDR depth, the H$_2$ formation rate (see Fig.\,\ref{fig:formation_H2}) is at least five times higher than the standard value for the diffuse ISM ($R_f\simeq 3-4\times 10^{-17}$ cm$^{3}$\,s$^{-1}$, see \cite{gry_h_2002}), which is what \cite{habart_empirical_2004} found from observations of H$_2$ lines in moderately excited PDRs ($G_0\simeq 10^{2}-10^{3}$).

It is also interesting to note that the difference between the position of the H~$\rightarrow$~H$_2$ transition for \qaf~= 17 \% (i.e. using diffuse ISM-like dust) to \qaf~= 2 \% (i.e. using Horsehead-like dust) is in the range of the spatial resolution of the JWST instruments ($\sim 0.1$\arcsec~to $\sim 1.2$\arcsec),
depending on the wavelength of the observed line (see Table \,\ref{tab:resolution}). Thus, the influence of the nano-grain depletion in the Horsehead on the position of the H~$\rightarrow$~H$_2$ transition should be observed with the JWST. One can note that the H~$\rightarrow$~H$_2$ transition also strongly depends on the density profile that must therefore be well constrained, which should be possible with the JWST.

\subsection{H$_2$ gas tracers}\label{sect:sect:gas_observables}

\begin{figure*}[h]
\centering
\includegraphics[width=0.57\textwidth, trim={0 0cm 0cm 0cm},clip]{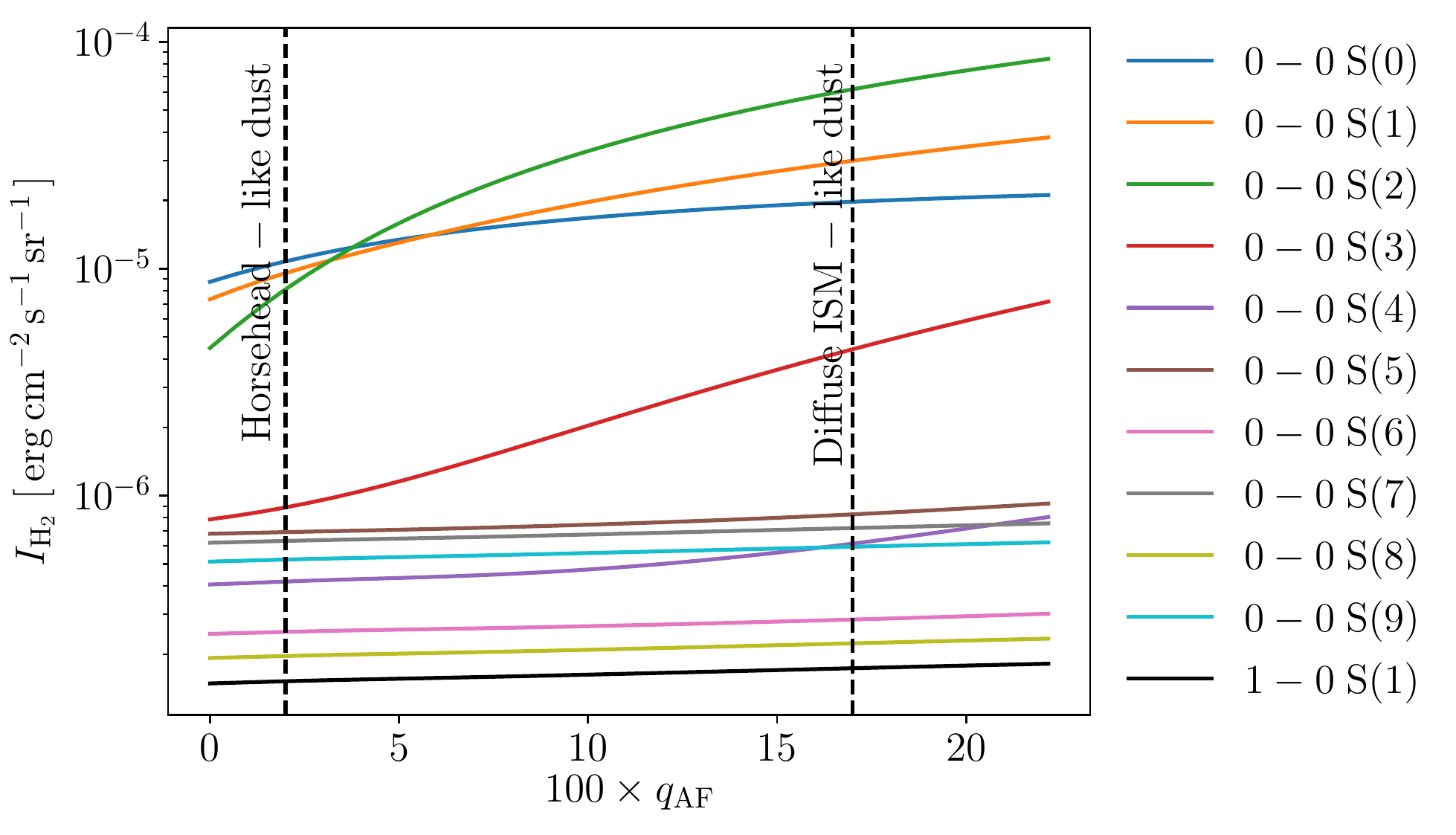}\hfill
\includegraphics[width=0.43\textwidth, trim={0 0cm 0cm 0cm},clip]{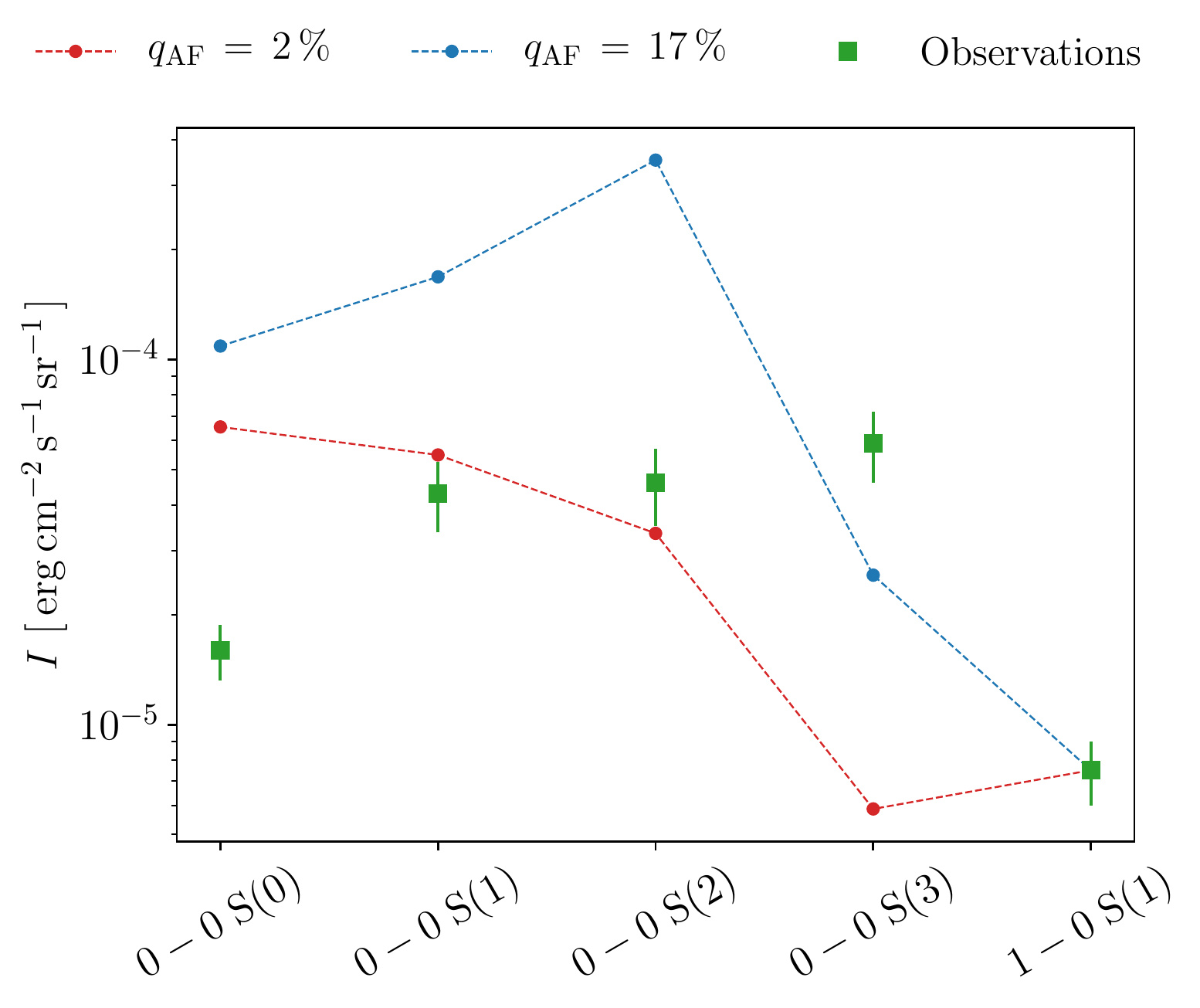}
    \caption{\textit{Left}: Modelled integrated intensities (i.e. intensity integrated along a line-of-sight considering a face-on PDR) of H$_2$ lines as a function of \qaf. \textit{Right}: Modelled integrated intensities of H$_2$ lines for \qaf~= 2 \% (red dots) and \qaf~= 17 \% (blue dots). Observed integrated intensity (The spectra are averaged in an area centred at the emission peak \citep[see black boxes in Fig.\,3 in][]{habart_excitation_2011}) in the Horsehead with uncertainties ($1\sigma$) is shown by the green squares. Models have been scaled on the 1-0 S(1) H$_2$ observed line.}
    \label{fig:intensities}
\end{figure*}

Variations in the gas temperature (see Sect.\,\ref{sect:sect:heating_source}) and in the chemistry (see Sect.\,\ref{sect:sect:chemistry}) affect the gas tracers. Figure \ref{fig:intensities} (left panel) shows the H$_2$ line integrated\footnote{The Meudon PDR code provides us with the integrated intensity along a line-of-sight across a PDR seen face-on.} intensities for \qaf~varying from 0 \% to 22 \%. We note that regardless of the considered line, the intensity of the first four pure rotational lines (e.g. 0-0 S(0), S(1), S(2), and S(3)) decreases with a decrease in \qaf. These lines essentially result from collisional excitation and therefore mainly depend on the gas temperature. Thus, as a decrease in \qaf~implies a decrease in the gas temperature (see Sect.\,\ref{sect:sect:heating_source}), these line intensities subsequently decrease with \qaf. 

The high rotational and ro-vibrational line 1-0 S(1) is essentially due to the decay of electronically states that are pumped through the absorption of FUV photons and the other pure rotational lines (e.g. 0-0 S(4), S(5), and up to S(9)) are essentially due to UV pumping \citep{habart_excitation_2011}. Moreover, the intensity of these lines is low compared to the first four pure rotational lines as the gas temperature is not high enough to pump them. Since there is not enough FUV and UV photons at the PDR depth of these lines and because this amount of photons barely depends on \qaf, these lines do not vary with \qaf. Decreases in the first four pure rotational lines with \qaf~are important: by a factor of $\sim 2$ for 0-0 S(0), a factor of $\sim 4$ for 0-0 S(1), a factor of $\sim 14$ for 0-0 S(2), and a factor of $\sim 6$ for 0-0 S(3) (see Fig.\,\ref{fig:intensities}, right panel). 

 In the right panel of Fig.\,\ref{fig:intensities}, we also compare the modelled and observed\footnote{The spectra are averaged in an area centred at the emission peak \citep[see black boxes in Fig.\,3 in][]{habart_excitation_2011}} intensities of the first four pure rotational lines 0-0 S(0) to S(3) and the ro-vibrational line 1-0 S(1), after normalising the last line because it barely depends on \qaf~and thus on the nano-grain depletion. This normalisation is justified as the modelled intensity is usually multiplied by a factor of $1/\cos(\theta)$ in order to take the geometry into account, which in the case of the Horsehead is $1/\cos(\theta)\sim 6.66$ \citep{habart_excitation_2011}. In our case, the normalisation of our two models requires two geometrical factors of 5.5 and 7.4, which are in agreement with the one used in \cite{habart_excitation_2011}.

We find good agreement, with \qaf~ = 2~\% for the 0-0 S(1) and 0-0 S(2) lines. The 0-0 S(3) line is underestimated by a factor of $\sim 10,$ at least, which is not surprising as this line is rarely reproduced by PDR models \citep[e.g.][]{habart_excitation_2011}, suggesting that understanding the excitation of H$_2$ and/or of heating processes in PDRs is still incomplete. The nano-grain depletion in PDRs is unfortunately going to decrease the gas temperature and, therefore, the 0-0 S(3) line.

\begin{table}[h]
    \centering
    \begin{tabular}{lcc}
        \hline
        \hline
        Line & $\lambda$ & R$_{\mathrm{JWST}}$ \\
         \hline
          0-0 S(0) & 28.2 $\mu$m  & 1.3\arcsec  \\
          0-0 S(1) & 17.0 $\mu$m  & 0.78\arcsec  \\
          0-0 S(2) & 12.3 $\mu$m  & 0.56\arcsec  \\
          0-0 S(3) & 9.7 $\mu$m   & 0.44\arcsec  \\
          1-0 S(1) & 2.1 $\mu$m   & 0.09\arcsec  \\
        \hline
    \end{tabular}
    \caption{Main H$_2$ lines with the associated intensities and the JWST spatial resolution at these wavelengths.}
    \label{tab:resolution}
\end{table}

\subsection{Influence of the dust extinction on the H$_2$ gas tracers}\label{sect:sect:assumption}

The column density at 0.025 pc is about $N_{\mathrm{H}}\sim 4.7\times 10^{20}$ H\,cm$^{-2}$, hence close but still below the critical value of $N_{\mathrm{H}}\sim 5\times 10^{20}$ H\,cm$^{-2}$ above which dust opacity becomes important \citep{hollenbach_photodissociation_1999}. This means that variations in the extinction in the outer irradiated part of the Horsehead barely affect the UV flux. Thus the gas heating the gas temperature therefore barely depends on these variations. Regarding the chemistry, the position of the H~$\rightarrow$~H$_2$ transition is dominated by the H$_2$ self-shielding if \G/$n_{\mathrm{H}}\la 4\times 10^{-2}$ cm$^{3}$ H$^{-1}$ and by dust extinction, conversely \citep{hollenbach_photodissociation_1999}. In this study, the H~$\rightarrow$~H$_2$ transition occurs between 0.010 and 0.015 pc (see Fig.\,\ref{fig:position_transition}), where the density is between $2500$ H\,cm$^{-3}$ and $7000$ H\,cm$^{-3}$ and \G/$n_{\mathrm{H}}$ is therefore between $1.4\times 10^{-2}$ cm$^{3}$ H$^{-1}$ and $4\times 10^{-2}$ cm$^{3}$ H$^{-1}$. Thus, the position of the H~$\rightarrow$~H$_2$ transition does not depend on the dust extinction, but on the H$_2$ self-shielding only. Therefore, variations in dust properties that could affect radiative transfer in the outer irradiated part of the Horsehead are negligible and the physical and chemical processes discussed in this paper would barely be affected if we were to take the nano-grain depletion into account in the radiative tranfer (i.e. $R_{\mathrm{V}}$ that varies). Thus, our assumption on a constant $R_{\mathrm{V}}$ is justified. 
One may note that if we had studied other mechanisms that occur in the denser part of the Horsehead such as the C$^{+}$~$\rightarrow$~C~$\rightarrow$~CO transition, this hypothesis would no longer have been valid to study this transition because, unlike the outer irradiated part of the Horsehead, variations in dust properties strongly affect the UV radiation field in the denser part of the Horsehead.    

\section{Summary and conclusion}\label{sect:summary_and_conclusion}

We study the influence of  nano-grain depletion observed in the outer irradiated region of the Horsehead on the gas physics and chemistry using the Meudon PDR code. As this PDR model does not include the THEMIS dust model that we used to highlight the nano-grain depletion, we had to tune the mass fraction of PAHs contained in the Meudon PDR code on a range that is, at first order, consistent with dust variations using THEMIS (i.e. the depletion of nano-grains is equivalent to a decrease in the mass fraction of PAHs, \qpah). We therefore calculated the value of \qpah~that corresponds to the dust properties derived in the Horsehead which differ from the diffuse ISM. We show the following.

\begin{enumerate}
    \item Gas heating is dominated by the photoelectric effect on nano-grains in the outer irradiated part of the Horsehead and because this dust population decreases when moving from a model with diffuse ISM-like dust (case \textit{a}) to a model with Horsehead-like dust (case \textit{b}), the gas temperature decreases as well from a factor of 1.5 to 3.
    \item The first four pure rotational lines of H$_2$ (e.g. 0-0 S(0), S(1), S(2), and S(3)) are very sensitive to a decrease in \qaf~as they are essentially excited through collisions and therefore depend on the gas temperature. The variations in these lines from case \textit{a} to \textit{b} range from a factor of 2 (for 0-0 S(0)) to a factor of 14 (0-0 S(2)).     \item The H~$\rightarrow$~H$_2$ transition shifts towards the inner part of the Horsehead because the H$_2$ formation decreases when moving from case \textit{a} to case \textit{b}. 
    \item The decrease in H$_2$ formation is due to the decrease in the total dust surface from case \textit{a} to case \textit{b}.
    \item The variation in the position of the H~$\rightarrow$~H$_2$ transition from case \textit{a} to case \textit{b} might be spatially resolved with the JWST, which was not the case with \textit{Spitzer}.
    \item The H$_2$ ro-vibrational line 1-0 S(1) is barely affected by variarions in \qaf~and therefore by the nano-grain depletion. 
\end{enumerate}

Even though the nano-grain depletion in the outer irradiated part of PDRs has been evidenced by observations for many years, the influence of this depletion on the gas tracers in PDRs has never been studied in detail. In this study, we quantitatively explore the influence of this depletion with the Meudon PDR code. This code does not include the THEMIS model, but  
thanks to a justified approximation we can mimic the nano-grain depletion with the dust model included in the Meudon PDR code.

We show that this depletion has a strong influence on the physics and the gas in the Horsehead. Moreover, the first four pure rotational lines of H$_2$ (e.g. 0-0 S(0), S(1), S(2), and S(3)) that are usually observed in PDRs can vary from a factor of 2 to 14. Also, the 0-0 S(3) line that is usually underestimated by models is underestimated even more with the nano-grain depletion due to the decrease in the heating through the photo-electric effect on dust grains. This clearly indicates that we need to consider another source of heating in the outer irratiated part of low excited PDRs such as the Horsehead. In order to go further, PDR models must include recent dust models that consider dust evolution from diffuse to dense regions. Given that current PDR models fail to fully explain the observed H$_2$ line intensities, in future work it would be worth exploring the contribution that other mechanisms could make. For example, the \cite{jones_h_2015} model proposes that the UV photon-processing of carbonaceous nano-particles forms H$_2$ and that it may operate under conditions where classical H$_2$ formation mechanisms are inhibited. The influence of this evolution on the gas tracers cannot be ignored if we want to advance our understanding of ISM physics and chemistry. 

\begin{acknowledgements} 
We thank the anonymous referee for very helpful sugges- tions and comments. We also thank Rachel Rudy, the language editor, for her correction. This work was supported by the Programme National “Physique et
Chimie du Milieu Interstellaire” (PCMI) of CNRS/INSU with INC/INP co-funded
by CEA and CNES. This work was also supported by the « P2IO LabEx (ANR-10-LABX-0038)»  in the framework "Investissements d'Avenir" (ANR-11-IDEX-0003-01) managed by the Agence Nationale de la Recherche (ANR), France.
\end{acknowledgements}

\bibliographystyle{aa} 
\bibliography{Zotero}

\begin{appendix}

\section{Influence of the cosmic ray ionisation rate on the H$_2$ gas tracers}
\label{appendix:cosmic_ray}

\begin{figure}[h]
\centering
\includegraphics[width=0.5\textwidth, trim={0 0cm 0cm 0cm},clip]{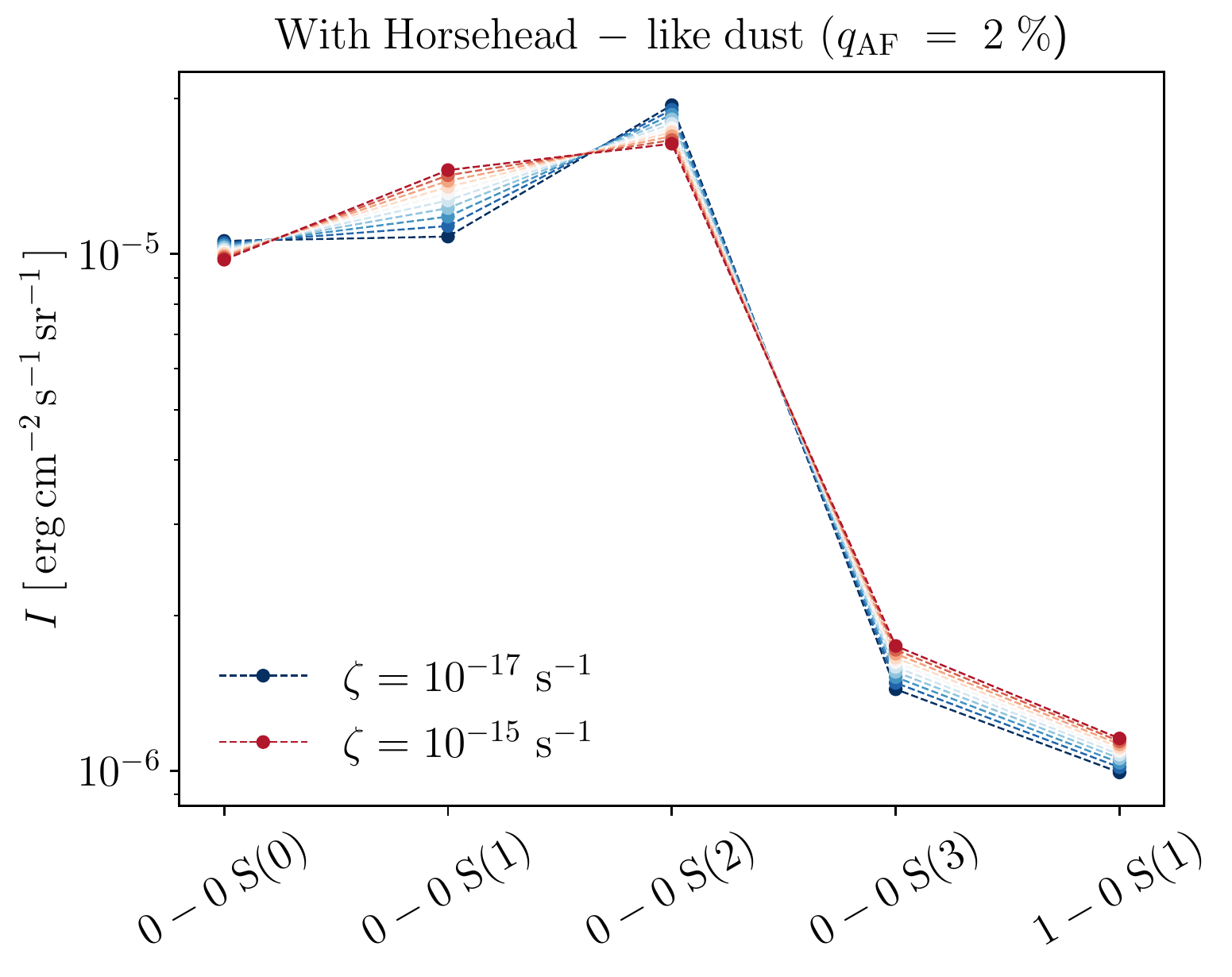}
    \caption{Modelled integrated intensities of H$_2$ lines for $\zeta$, varying from $10^{-17}$ s$^{-1}$ to $10^{-15}$ s$^{-1}$ using Horsehead-like dust (i.e. \qaf~= 2 $\%$).}
    \label{fig:zeta_HH}
\end{figure}

\begin{figure}[h]
\centering
\includegraphics[width=0.5\textwidth, trim={0 0cm 0cm 0cm},clip]{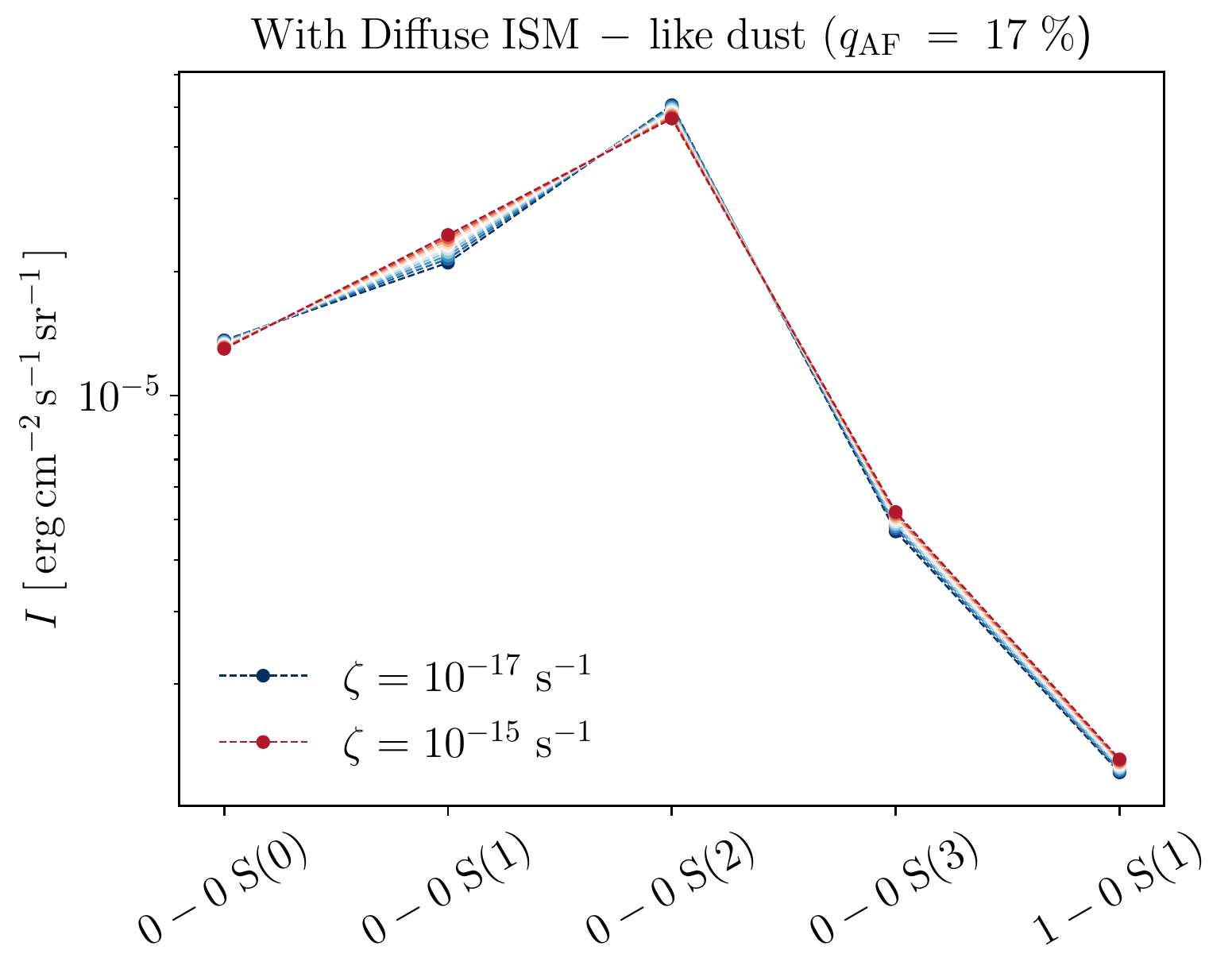}
    \caption{Modelled integrated intensities of H$_2$ lines for $\zeta$, varying from $10^{-17}$ s$^{-1}$ to $10^{-15}$ s$^{-1}$ using diffuse ISM-like dust (i.e. \qaf~= 17 $\%$).}
    \label{fig:zeta_diffuse}
\end{figure}

Here, we study the influence of the cosmic ray ionisation rate on the H$_2$ gas tracers. Fig.\,\ref{fig:zeta_HH} shows the modelled integrated intensities of H$_2$ lines for $\zeta$ varying from $10^{-17}$ s$^{-1}$ to $10^{-15}$ s$^{-1}$ using Horsehead-like dust (i.e. \qaf~= 2 $\%$) and Fig.\,\ref{fig:zeta_diffuse} shows the same thing but with diffuse ISM-like dust (i.e. \qaf~= 17 $\%$). Regardless of the dust used, the gas tracers barely vary (less than 1 $\%$ of variations) with $\zeta$, hence the cosmic rays do not influence the H$_2$ gas tracers.

\section{Number density of H$_2$ for the different lines}

Here, we show the number density of H$_2$ for the different lines as a function of the depth inside the Horsehead for \qaf~varying from 0 to 22 $\%$.

\begin{figure}[h]
\centering
\includegraphics[width=0.5\textwidth, trim={0 0cm 0cm 0cm},clip]{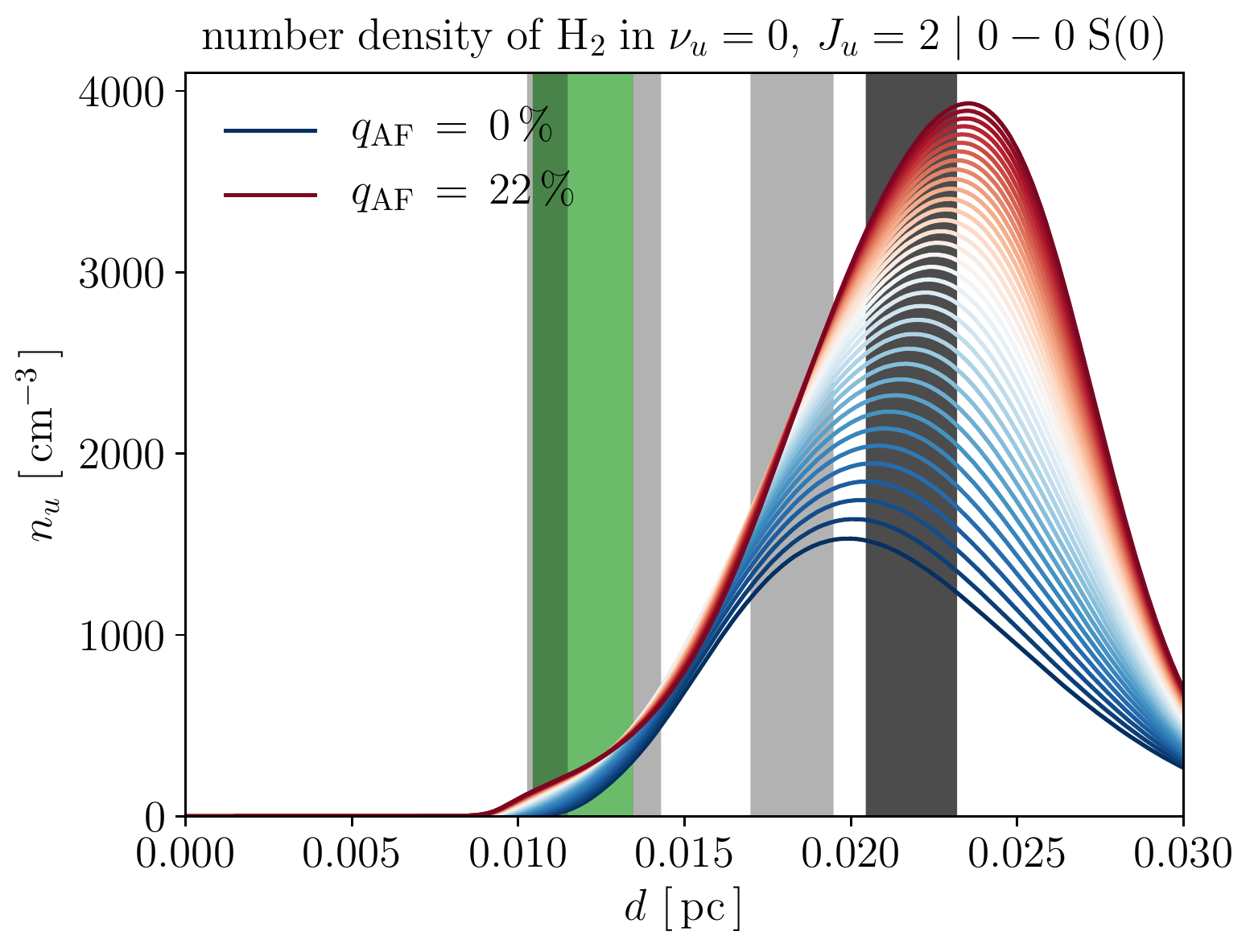}
    \caption{Number density of H$_2$ in the state $\nu_u=0$ and $J_u=2$ across the Horsehead. The meaning of the green and grey stripes is explained in Fig.\,\ref{fig:temp_q_pah}. The darker grey stripe corresponds to the position of the maximum of this line for \qaf~varying from 2 \% to 17 \%.}
    \label{fig:0_0_S_0}
\end{figure}

\begin{figure}[h]
\centering
\includegraphics[width=0.5\textwidth, trim={0 0cm 0cm 0cm},clip]{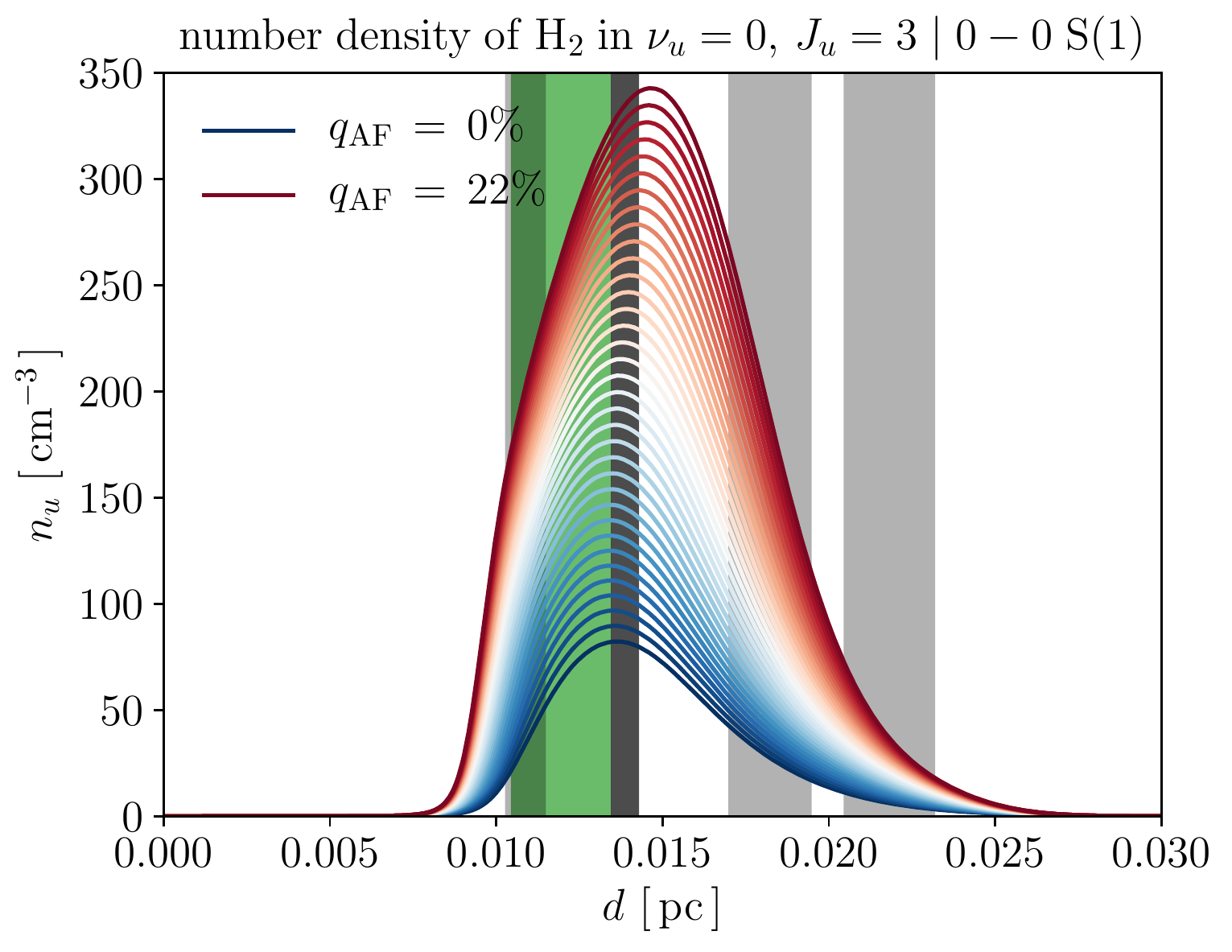}
    \caption{Number density of H$_2$ in the state $\nu_u=0$ and $J_u=3$ across the Horsehead. The meaning of the green and grey stripes is explained in Fig.\,\ref{fig:temp_q_pah}. The darker grey stripe corresponds to the position of the maximum of this line for \qaf~varying from 2 \% to 17 \%.}
    \label{fig:0_0_S_1}
\end{figure}

\begin{figure}[h]
\centering
\includegraphics[width=0.5\textwidth, trim={0 0cm 0cm 0cm},clip]{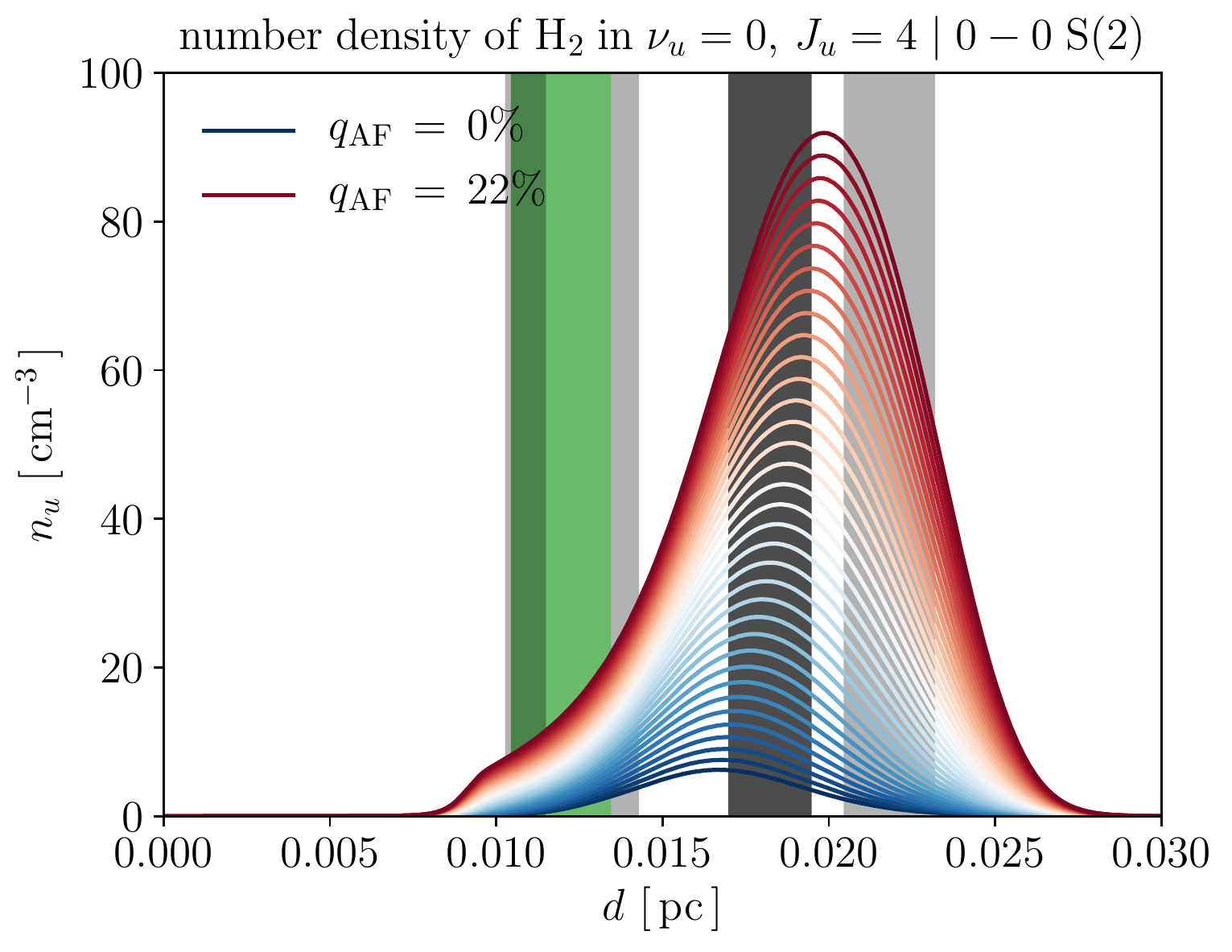}
    \caption{Number density of H$_2$ in the state $\nu_u=0$ and $J_u=4$ across the Horsehead. The meaning of the green and grey stripes is explained in Fig.\,\ref{fig:temp_q_pah}. The darker grey stripe corresponds to the position of the maximum of this line for \qaf~varying from 2 \% to 17 \%.}
    \label{fig:0_0_S_2}
\end{figure}

\begin{figure}[h]
\centering
\includegraphics[width=0.5\textwidth, trim={0 0cm 0cm 0cm},clip]{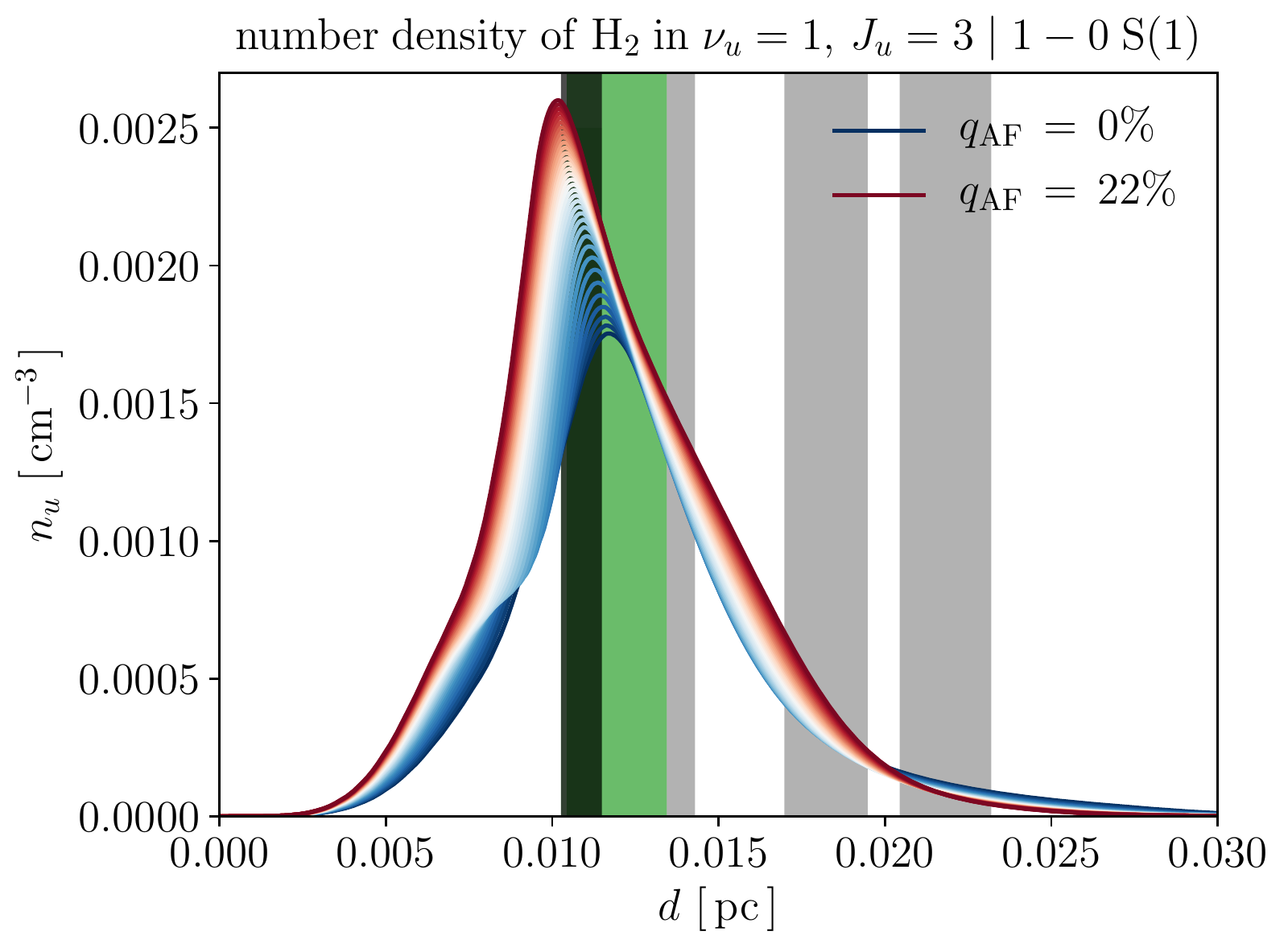}
    \caption{Number density of H$_2$ in the state $\nu_u=1$ and $J_u=3$ across the Horsehead. The meaning of the green and grey stripes is explained in Fig.\,\ref{fig:temp_q_pah}. The darker grey stripe corresponds to the position of the maximum of this line for \qaf~varying from 2 \% to 17 \%.}
    \label{fig:1_0_S_1}
\end{figure}

\end{appendix}

%
%

\end{document}